\documentclass[aps,pra,twocolumn,a4paper,showpacs]{revtex4-1}
 \usepackage{latexsym}
 \usepackage{amsmath}
 \usepackage{amsfonts}
 \usepackage{graphicx}
 \usepackage{amssymb}
 \usepackage{subfigure}
 \usepackage{color}
 \newcommand{\ket}[1]{\ensuremath{|#1\rangle}}
 \newcommand{\bra}[1]{\ensuremath{\langle #1 |}}

 \newcommand{\bc}{\begin{center}}
 \newcommand{\ec}{\end{center}}

 \newcommand{\ii}{i}
 \newcommand{\DC}{\Delta_{_{1}}}
 \newcommand{\DP}{\Delta_{_{2}}}
 \newcommand{\DM}{\Delta_{_{3}}}
  \newcommand{\DR}{\Delta_{_{R}}}

\begin{document}
\title{Steering, Splitting and Cloning of Optical Beam in a Coherently Driven Raman Gain System}
\author{Onkar N. \surname{Verma}}
\email{onkar@iitg.ernet.in}
\author{Tarak N. \surname{Dey}}
\email{tarak.dey@iitg.ernet.in}
\affiliation{Department of Physics, Indian Institute of Technology Guwahati, Guwahati- 781 039, Assam, India}
\begin{abstract}
We propose an all-optical anti-waveguide mechanism for steering, splitting, and cloning of an optical beam beyond the diffraction-limit. 
We use a spatially inhomogeneous pump beam to create an anti-waveguide structure in a Doppler broadened $\mathcal{N}$-type four-level Raman gain medium for a co-propagating weak probe beam. 
We show that a transverse modulated index of refraction and gain due to the spatially dependent pump beam hold the keys to steering, splitting and cloning of an optical beam. 
We have also shown that an additional control field permits the propagation of an optical beam through an otherwise gain medium without diffraction and instability. 
We further discuss how finesse of the cloned images can be increased by changing the detuning of the control field. 
\end{abstract}
\maketitle
\section{Introduction} 
Optical beam guiding, deflection and cloning has attracted a great deal of attention due to its tremendous applications in optical imaging, optical switching, optical lithography, laser machining,  and free-space communication technologies.
The guiding and steering of an optical beam is made possible by virtue of a refractive index of the medium.
Several techniques such as mechanical~\cite{I_Cindrich,D_H_McMahon}, thermal ~\cite{W_B_Jackson}, electrical~\cite{J_D_Zook},  acousto-optical~\cite{R_W_Dixon}  and all optical  ~\cite{g_p_agrawal,a_t_ryan,y_li,a_j_stentz} have been proposed to control the refractive index for beam deflection.
However,  all-optical methods have been paid much effort  owing to many striking features such as high speed, efficiency, and quick nonlinear response time.

The nonlinear optical interactions between light and matter creates a new avenue to control over beam propagation dynamics through a medium.
This is feasible as the absorptive and dispersive properties of the medium can be modified by the strength of the interactions.
Such manipulation of dispersion and absorption leads to many novel phenomena including electromagnetically induced transparency (EIT)~\cite{harris,michael}, 
coherent population trapping(CPT)~\cite{e_arimondo}, saturated absorption techniques~\cite{t_w_hansch,g_s_agarwal_09} or lasing without inversion(LWI)~\cite{olga_kocha}.  
The sharp refractive index changes near the centre of the transparency window for the EIT medium is the key concept for beam deflection~\cite{sautenkov,zhou,sun}.
The ability to control of light deflection is also possible by use of transverse magnetic field through an atomic medium~\cite{schlesser,holzner,Karpa_Nat_06}.
Further a suitable spatially dependent control field can be used to modulate the refractive index along the transverse direction.
This spatially modulated refractive index generates several effects such as induced focusing~\cite{Moseley1,Moseley2,focusing3,focusing4,mitsunaga}, 
waveguiding~\cite{truscott,kapoor,andersen,mukund} and anti-waveguiding~\cite{defocusing}.  

Most of the EIT-based schemes for producing beam deflection and  guiding have low transmission due to presence of medium absorption~\cite{sautenkov,zhou,sun}.
Therefore, finding an alternative medium which displays gain with the desired variation of refractive index is a challenging task.
In this context, active Raman gain(ARG) media have attracted a lot of attention~\cite{agarwal}.
Recently Zhu~{\it et al.}~\cite{C_Zhu_13} have theoretically studied the beam deflection in an ARG medium. 
They have used spatially inhomogeneous pump beam to deflect a weak probe beam. 
They have found that the deflection angle is increased by an order of magnitude as compared to EIT medium. 
Nonetheless the probe field experiences a large amount of gain during the propagation through a $\Lambda$-type ARG medium~\cite{C_Zhu_13,agarwal}.
This large gain makes the probe beam propagation unstable and thus limits the practical application~\cite{agrawal1,agrawal2,Gisin}.
Moreover, the input spot sizes for individual Gaussian profiles of pump and probe beams are equal to $1.4$ cm and $1$ mm, respectively. 
Hence, the diffraction spreading of such beams are not relevant since Rayleigh length is much larger than the length of the medium. 
Focusing laser beams into smaller spots~\cite{dey} and increasing the spatial resolution of arbitrary images~\cite{om2} 
is a fundamental problem in all-optical image processing~\cite{Lantz_Nat_08,Cohen_Nat_10}.
Distortion and absorption holds the fundamental limitation for the creation, detection, or propagation of small images.
This limitation affects the applications such as efficient transfer and conversion of small images~\cite{Li,om1,ding,cao}, steering~\cite{wang,Hang,lida1} or optical manipulation of light beams~\cite{lida2}.
Here we address these issues by considering the propagation of diffraction-limited beams and arbitrary images through a controllable ARG medium.

In this paper we exploit an anti-waveguide mechanism~\cite{defocusing} to show beam steering, splitting, and cloning of an arbitrary images in an inhomogeneously broadened medium.
To facilitate these processes, we use spatial inhomogeneous pump beam to write an anti-waveguide inside the medium for co-propagating probe beam.
At two-photon Raman detuning condition, the refractive index and gain of the probe susceptibility are high at the peak of the Gaussian pump beam whereas at wings both are very small.
The high refractive index together with gain allow to deflect the probe beam when it is launched at the wings of the pump beam.
The control field parameters such as detuning and intensity can be used to control the transmission intensity and width of the deflected probe beam.
Next, we reveal splitting of a single super-Gaussian probe beam into two Gaussian beams by use of two-peak pump beam structure.
The bright(cladding) and dark(core) regions of the pump field profile induces a high(cladding) and low(core) refractive index of the probe field which lead to formation of an anti-waveguide structures inside the medium.   
More specifically,  the super-Gaussian probe beam guided out from the core where it was injected.
The diffraction-limited probe beam gets focus in the cladding due to the converging refractive index.
We also observe that the transmitted probe beam gets the shape of the pump beam with finesse two times larger than the initial finesse of the pump beam.
Further, we  demonstrate the cloning of a doughnut-shaped pump beam structure onto the probe beam. 
Our numerical simulation shows that the cloned probe has a controllable gain with high finesse. 
Furthermore, our scheme can be employed for cloning the arbitrary pump images to the probe beam even though the pump images are severely distorted due to diffraction.
It follows that our findings can greatly improve the device performance on beam steering, splitting and image cloning beyond the diffraction limit.

The article is organized as follows.
In the next section, we introduce the physical model and basic equations of motion for a four-level system. 
In Sec. III, an approximate expression for a linear susceptibility of a weak probe field is derived using perturbative approach. 
We include the thermal motion of the atoms by averaging the susceptibility over Maxwell-Boltzmann velocity distribution.
In Sec. IV, we describe the beam propagation equations for the evolution of both pump and probe fields under paraxial approximations. 
In Sec. V, we discuss our results based on numerical simulation. 
We first explain the spatially dependent susceptibility for different shapes of the pump beam and advantage of a uniform control beam.
We then perform numerical integration of the beam propagation equations in order to demonstrate steering, splitting and cloning of an optical beam.
Obtained results are summarised in the final section.
\begin{figure}[t!]
\includegraphics[width=1.\columnwidth]{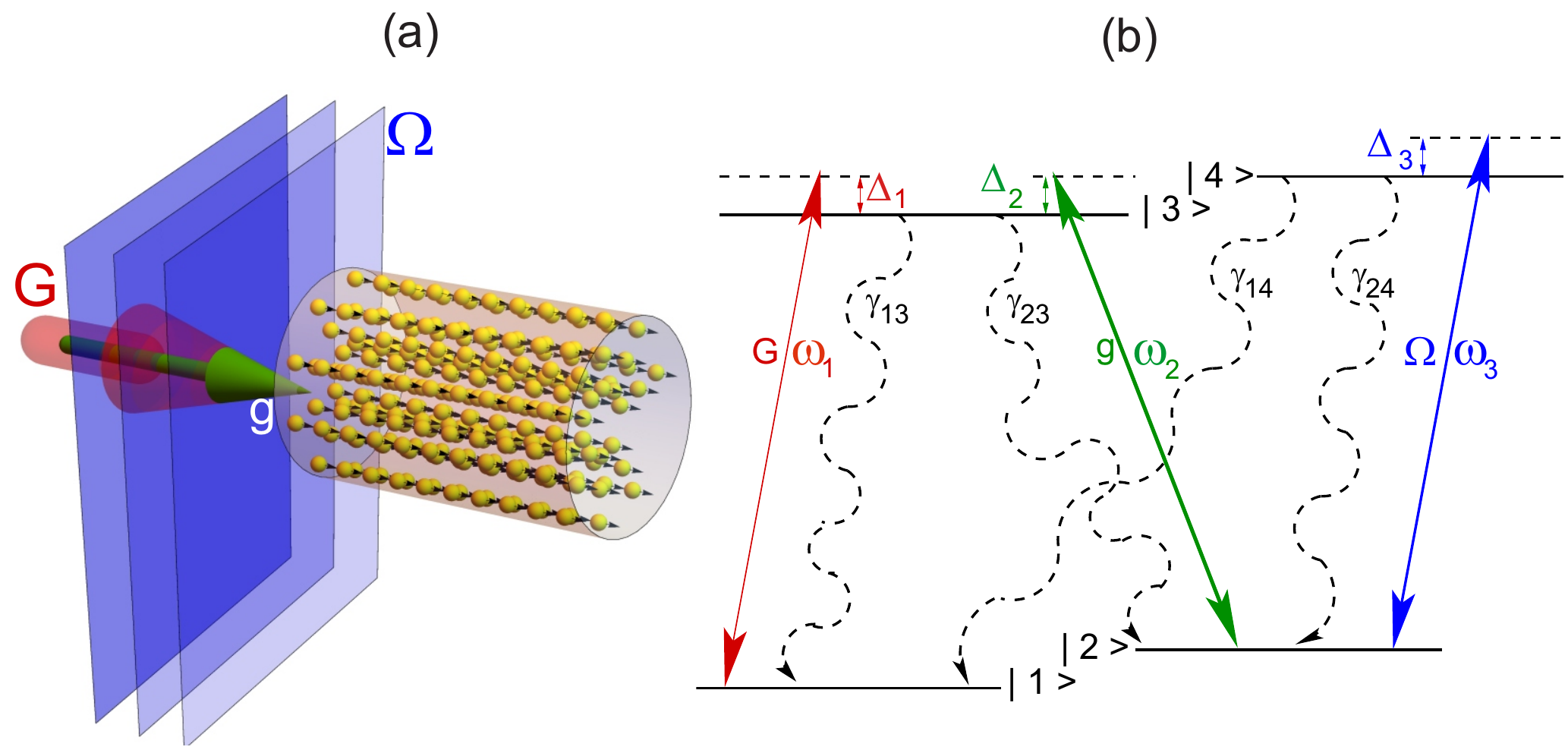}
\caption{\label{fig:Fig1} (Color online) 
\textcolor{black}{(a)Schematic illustration to produce steering, splitting and cloning of the optical beam.
The beam shaped pump, probe and a plane wave control fields are co-propagating with the thermal $^{87}$Rb atoms. 
(b)Energy-level diagram of a four-level $^{87}$Rb atomic system in $\mathcal{N}$ configuration.  
The atomic transition $\ket{3}\leftrightarrow\ket{1}$ is coupled by a pump field of Rabi frequency $G$. 
The weak probe field of Rabi frequency $g$ interacts with the atomic transition $\ket{3}\leftrightarrow\ket{2}$. 
A control field of Rabi frequency $\Omega$ connects the transition $\ket{4}\leftrightarrow\ket{2}$ to produce controllable gain of the system.}}
\end{figure}
\section{Physical Model and Basic Equations}
The schematic of the system under consideration for the generation of steering, splitting and cloning of an optical beam is illustrated in Fig.\ref{fig:Fig1}(a) where
three co-propagating fields interact within the inhomogeneously broadened medium.
The electrical dipole allowed transitions $\ket{1}\leftrightarrow\ket{3}$,   $\ket{3}\leftrightarrow\ket{2}$, and  $\ket{2}\leftrightarrow\ket{4}$ form a four-level $\mathcal{N}$-type  atomic system as shown in Fig.\ref{fig:Fig1}(b).
The transitions $\ket{1}\leftrightarrow\ket{2}$, $\ket{3}\leftrightarrow\ket{4}$ and $\ket{1}\leftrightarrow\ket{4}$ are generally forbidden electric dipole transitions.
The atomic transitions $\ket{3}\leftrightarrow\ket{1}$,   $\ket{3}\leftrightarrow\ket{2}$, and  $\ket{4}\leftrightarrow\ket{2}$ are driven by
a pump field with frequency $\omega_{_{1}}$, a weak probe field with frequency $\omega_{_{2}}$ and a control field with frequency $\omega_{_{3}}$, respectively. 
This generic level configuration can be found for example in energy levels of $^{87}$Rb which contain ground levels $\ket{1}=\ket{5S_{1/2}, F=2}$, $\ket{2}=\ket{5S_{1/2},F=3}$  and excited levels $\ket{3}=\ket{5P_{1/2},F'=2}$ and $\ket{4}=\ket{5P_{3/2},F'=4}$, respectively~\cite{min_yan,Li_prl}.

We define three co-propagating electric fields as follows:
\begin{align}\label{field}
 {\vec{E}_j}(\vec{r},t)= \hat{e}_{j}\mathcal{E}_{j}(\vec{r})~e^{- i\left(\omega_j t-  k_j z\right )} + {c.c.},
\end{align}
where, $\mathcal{E}_{j}(\vec{r})$ are slowly varying envelopes, $\hat{e}_{j}$ is the unit polarization vector, and $k_j$ is the wave number of electric fields. 
The index $j\in \{1,2,3\}$ denotes the pump, probe, and control fields, respectively.
Under the action of three coherent fields, the interaction Hamiltonian of the system in the dipole and rotating wave approximation is given by
\begin{align}
\label{Heff}
& {\mathcal{H}_I}/{\hbar}=(\DP-\DC-\DM)\ket{4}\bra{4}-(\DC-\DP)\ket{2}\bra{2}    \notag \\
 &- \DC\ket{3}\bra{3} - (g\ket{3}\bra{2} + G \,\ket{3}\bra{1} \,+ \Omega \,\ket{4}\bra{2} + \text{H.c.})\,,
\end{align}
where $\DC = \omega_{_{1}} - \omega_{_{31}}$, $\DP = \omega_{_{2}} - \omega_{_{32}}$, $\DM = \omega_{_{3}} - \omega_{_{42}}$ 
are the single-photon detunings of the pump, probe, and control fields, respectively.
The atomic transition frequencies are denoted by $\omega_{ij}$.
The Rabi frequencies of pump, probe and control fields are defined as 
\begin{align}
\label{Rabi_frequencies}
G=\frac{\vec{d}_{13}\cdot\vec{\mathcal{E}}_{\rm{1}}}{\hbar},~~g=\frac{\vec{d}_{23}\cdot\vec{\mathcal{E}}_{\rm{2}}}{\hbar}, ~{\textrm {and}} ~~\Omega=\frac{\vec{d}_{24}\cdot\vec{\mathcal{E}}_{\rm{3}}}{\hbar}\,,
\end{align}
where the $d_{ij}$ are the corresponding dipole moment matrix elements of transitions $\ket{i}\leftrightarrow\ket{j}$.

The dynamical evolution of the atomic system can be described by the density matrix equations~\cite{om2}, 
\begin{align}
\label{master}
\dot{\rho}=-\frac{\ii}{\hbar}\left[\mathcal{H}_I,\rho\right]+\mathcal{L}\rho\,.
\end{align}
where the Liouvillian matrix $\mathcal{L}\rho$, defined in Eq. (\ref{Liouvillian}), describes the relaxation by radiative and non-radiative decay
\begin{widetext}
\begin{align}
\label{Liouvillian}
\mathcal{L}\rho = \left[ \begin{array}{cccc}
\gamma_{_{13}}\rho_{_{33}}+\gamma_{_{14}}\rho_{_{44}}&-\gamma_{_{c}}\rho_{_{12}}& -\Gamma_{_{13}}\rho_{_{13}}& -\Gamma_{_{14}}\rho_{_{14}}\\
-\gamma_{_{c}}\rho_{_{21}}&\gamma_{_{23}}\rho_{_{33}}+\gamma_{_{24}}\rho_{_{44}}&-\Gamma_{_{23}}\rho_{_{23}}&-\Gamma_{_{24}}\rho_{_{24}}\\
 -\Gamma_{_{31}}\rho_{_{31}}&-\Gamma_{_{32}}\rho_{_{32}}&-(\gamma_{_{13}}+\gamma_{_{23}})\rho_{_{33}} &-\Gamma_{_{34}}\rho_{_{34}}\\
-\Gamma_{_{41}}\rho_{_{41}}&-\Gamma_{_{42}}\rho_{_{42}}&-\Gamma_{_{43}}\rho_{_{43}}&-(\gamma_{_{14}}+\gamma_{_{24}})\rho_{_{44}}\\\end{array} \right].
 \end{align}
 \end{widetext}
The radiative decay rates from the excited states $|3\rangle$ and $|4\rangle$ to ground states $|1\rangle$ 
and $|2\rangle$ are labeled by $\gamma_{_{i3}}$ and $\gamma_{_{i4}}$, $i\in \{1,2\}$ and 
the collisions dephasing rate $\gamma_{_{c}}$ describes redistribution of populations between ground levels.
The decay rate of the atomic coherence is defined as
\begin{align}
\label{atomic_coherence_decay}
\Gamma_{_{\alpha\beta}}=\frac{1}{2}\left[\displaystyle\sum_{i}\gamma_{_{i\alpha}}+\displaystyle\sum_{i}\gamma_{_{i\beta}}\right]+\gamma_{_{c}},~i\notin \{\alpha,\beta\}\,.
\end{align}
Substituting the interaction Hamiltonian of Eq.(\ref{Heff}) and the Liouvillian matrix of Eq.(\ref{Liouvillian}) in the density matrix  Eq.(\ref{master}), 
the equations of motion for the four-level atomic system can be described as
\begin{align}
\label{Full_density}
 \dot{\rho}_{_{11}}&=\gamma_{_{13}}\rho_{_{33}} +\gamma_{_{14}}\rho_{_{44}} + {\ii} G^* \rho_{_{31}} -  {\ii} G \rho_{_{13}}  \,,\nonumber\\
 \dot{\rho}_{_{22}}&=\gamma_{_{23}}\rho_{_{33}} + \gamma_{_{24}}\rho_{_{44}} + {\ii} g^* \rho_{_{32}} -  {\ii} g \rho_{_{23}} + {\ii} \Omega^* \rho_{_{42}}-  {\ii} \Omega \rho_{_{24}} \,,\nonumber\\
  \dot{\rho}_{_{21}}&=-\left[ \gamma_{c} - \ii \DR \right]\rho_{_{21}} - {\ii} G \rho_{_{23}} + {\ii} g^{*} \rho_{_{31}} + {\ii} \Omega^{*} \rho_{_{41}}\,,\nonumber\\
\dot{\rho}_{_{33}}&=-(\gamma_{_{13}}+\gamma_{_{23}})\rho_{_{33}} + {\ii} G \rho_{_{13}} -  {\ii}  G^* \rho_{_{31}} +  {\ii}  g\rho_{_{23}} -  {\ii} g^*\rho_{_{32}}\,,\nonumber\\
 \dot{\rho}_{_{31}}&=-\left[\Gamma_{_{31}} - \ii \DC\right]\rho_{_{31}} + {\ii} g \rho_{_{21}} + {\ii} G (\rho_{_{11}}- \rho_{_{33}}) \,,\nonumber\\
 \dot{\rho}_{_{32}}&=-\left[\Gamma_{_{32}} - \ii \DP\right]\rho_{_{32}} + {\ii} G\rho_{_{12}} - {\ii} \Omega \rho_{_{34}} + {\ii} g( \rho_{_{22} }- \rho_{_{33}})\,,\nonumber\\
 \dot{\rho}_{_{34}}&=-\left[ \Gamma_{_{34}} - \ii (\DP-\DM)\right]\rho_{_{34}} + {\ii} G \rho_{_{14}}+ {\ii} g \rho_{_{24}} - {\ii} \Omega^*\rho_{_{32}} \,,\nonumber\\
 \dot{\rho}_{_{41}}&=-\left[ \Gamma_{_{41}}- \ii (\DR+\DM)\right]\rho_{_{41}} + \ii \Omega \rho_{_{21}}- {\ii} G\rho_{_{43}} \,,\nonumber\\ 
 \dot{\rho}_{_{42}}&=-\left[ \Gamma_{_{42}}- \ii \DM\right]\rho_{_{42}} + {\ii} \Omega (\rho_{_{22}} -  \rho_{_{44}})- {\ii} g \rho_{_{43}} \,,
 \end{align}
together with population conservation condition $ \rho_{_{11}}+\rho_{_{22}}+\rho_{_{33}}+\rho_{_{44}}=1$ and two-photon Raman detuning $\DR=\DC-\DP$.
In the next section, we obtain the analytical expression for the linear susceptibility of the probe field in a compact form with the assumption of equal
decay rates from excited states, {\it {i.e.,}}
$\gamma_{_{13}}=\gamma_{_{23}}=\gamma_{_{14}}=\gamma_{_{24}}=\gamma/2$ .
\section{ Probe Susceptibility for Hot atomic medium}
In this section, we derive an approximate solution of linear susceptibility of the probe field in a hot atomic medium.
The analytical solution of the atomic coherence ${\rho}_{_{32}}$ for the probe field can be obtained by solving the density matrix Eqs.(\ref{Full_density}) in the steady state condition. 

We assume that all atoms are prepared initially in the ground state $|1\rangle$.
Due to the presence of large detuning of the strong pump and weak probe fields, most of the atoms populate at their ground state $|1\rangle$ while other states $|j\rangle (j\neq1)$, remain empty at later time.
Hence the system turns to an ARG configuration for the probe field. 
Since Raman gain process is basically a second-order process,
we therefore expand the density matrix elements to first order in the probe field $g$ and to second order in the pump field $G$ but all orders in the control field $\Omega$ in the weak probe field limit. 
The perturbation expansion of the density matrix can be expressed as
\begin{align}
\rho_{_{ij}}&= \rho_{_{ij}}^{(0)} + G\rho_{_{ij}}^{(1)} + G^*\rho_{_{ij}}^{(2)} + g\rho_{_{ij}}^{(3)} + g^*\rho_{_{ij}}^{(4)} + G^2\rho_{_{ij}}^{(5)}\nonumber\\
              & + |G|^2\rho_{_{ij}}^{(6)} + {G^*}^2\rho_{_{ij}}^{(7)}  + gG\rho_{_{ij}}^{(8)} + gG^*\rho_{_{ij}}^{(9)} \nonumber\\
              & + g^*G\rho_{_{ij}}^{(10)} + g^*G^*\rho_{_{ij}}^{(11)} + g|G|^2\rho_{_{ij}}^{(12)},
\end{align}
where, $\rho_{_{ij}}^{(0)}$ describes the solution in the absence of all three optical fields and  $\rho_{_{ij}}^{(k)}$ denotes the $k$-th order solution. 
Now we substitute the above expression in the Eqs.(\ref{Full_density}) and equate the coefficients of $g$, $g^{*}$, $G^n$ ($n\in1,2$), and constant terms. 
As a result, we obtain a set of 12 coupled simultaneous linear algebraic equations to determine the expression of ${\rho}_{_{32}}^{(12)}$.
We use back substitution method to solve these algebraic equations in order to derive the probe coherence ${\rho}_{_{32}}$.
The different terms in the expression of the probe coherence are given in Appendix~\ref{app-A}.
The atomic coherence ${\rho}_{_{32}}$ will yield the probe susceptibility $\chi$ at frequency $\omega_{_{2}}$
\begin{equation}\label{chi_32}
{\chi}(\DP)=\frac{N|d_{_{32}}|^2}{{\hbar}}{\rho}_{_{32}}\,,
\end{equation}
where $N$ is the atomic density of the medium. 
The above analysis is valid for stationary atoms.
While for a hot atomic system, the thermal motion of the atoms causes inhomogeneous broadening of the atomic spectra.
The thermal velocity $v$ of the atom can be included in the susceptibility expression(\ref{chi_32}) by introducing velocity-dependent field detunings 
$\Delta_{j}(v)=\Delta_{j}-k_{j}v,~j\in \{1,2,3\}$. 
The term $k_{j}v$ is the Doppler shift experienced by an atom with a velocity component $v$ in the direction of the beam propagation of the fields.
We have assumed the wave vectors of the three fields are nearly equal ($k_j\approx k$).
The negative sign in the velocity-dependent field detuning $\Delta_{j}(v)$ indicate that atom and field are co-propagating.
The susceptibility of a hot atomic vapour system needs to be averaged over the entire velocity distribution of atoms and it is given by
\begin{equation}\label{chi_average}
\langle\chi\rangle=\int_{-\infty}^{\infty} \chi(kv) P(kv)d(kv)\,.
\end{equation}
The velocity distribution of the atom is assumed to obey the Maxwell-Boltzmann distribution
\begin{equation}
\label{MB}
P(kv)d(kv)=\frac{1}{\sqrt{2\pi \mathcal{D}^2}}e^{-\frac{(kv)^2}{2{\mathcal{D}}^2}}d(kv)\,.
\end{equation}
The Doppler width $\mathcal{D}$ at temperature $T$ defined by
\begin{equation}
\label{Doppler_width}
\mathcal{D}=\sqrt{\frac{k_{B}T\omega^2}{Mc^2}}\,, 
\end{equation}
where $M$ is the atomic mass and $k_B$ is the Boltzmann constant.
Doppler broadening plays a crucial role to control the width of the absorption or gain window of the thermal media~\cite{Agarwal,Kash,Javan,Peng}.
The spectral features of window become narrower in a Doppler broadened medium as compared with the homogeneous medium. 
The steepness of the refractive index due to the narrowing of resonance window can be useful in many applications such as slow light, storage of light and high resolution spectroscopy. 
Thus we include atomic velocity effect on the beam propagation dynamics through ARG medium by considering Doppler averaging in the susceptibility expression.
\section{Beam propagation equations and beam profiles}
\begin{figure}[t!]
\includegraphics[width=1.\columnwidth]{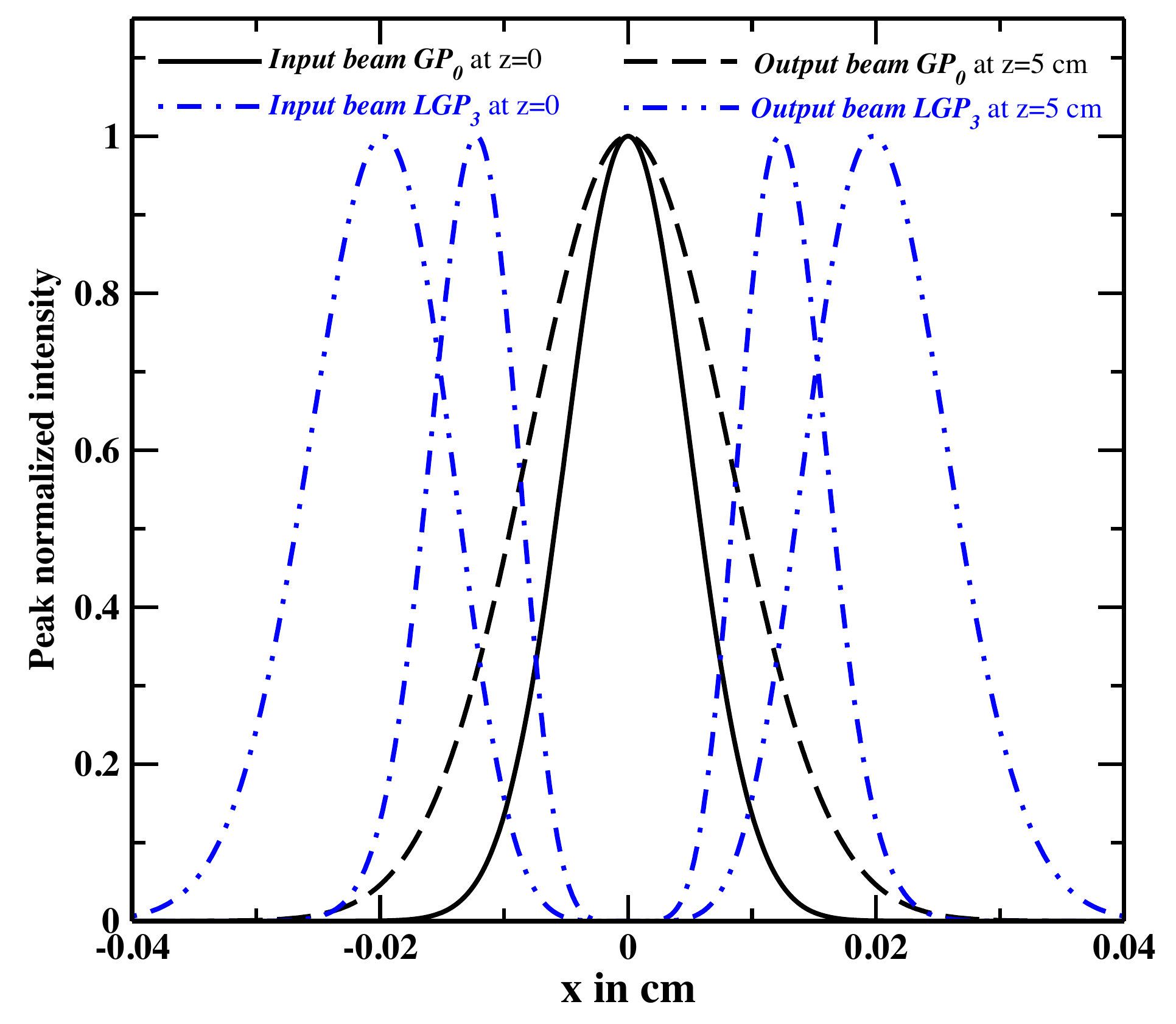}
\caption{\label{fig:Figure2} (Color online) 
\textcolor{black}{ Pump  intensities profile for two different shapes namely Gaussian (GP$_0$) and Laguerre-Gaussian(LGP$_3$) is plotted against $x$ at $y=0$ plane.
The initial amplitude and width of profiles are $G_{0}=2\gamma$ and ${w_0}=100~{\mu}$m, respectively.}}
\end{figure}

The propagation of co-propagating pump and probe fields with amplitudes $\mathcal{E}_1$ and $\mathcal{E}_2$ along the $z$-direction are governed by Maxwell's wave equations.
Under slowly varying envelope and paraxial wave approximations, 
the beam propagation equations for pump and probe field can be expressed in the following form
\begin{subequations}
\begin{align}
 \frac{\partial G}{\partial z}
   &= \frac{i}{2{k_1}} \left( \frac{\partial^2 }{\partial x^2}
      + \frac{\partial^2 }{\partial y^2} \right) G \,,\label{beam_eq_control} \\
 \frac{\partial g}{\partial z}
   &= \frac{\ii }{2{k_2}} \left( \frac{\partial^2 }{\partial x^2}
      + \frac{\partial^2 }{\partial y^2} \right) g + 2i{\pi}k_1\langle{\chi}\rangle{g}\label{beam_eq_probe} \,. 
\end{align}
\end{subequations}
The velocity-averaged susceptibility $\langle\chi\rangle$ is included only in the probe beam equation, whereas this effect is very negligible on the pump beam propagation under the weak probe field~\cite{om1}.
The second partial derivatives in the transverse directions $(x,y)$ represent a paraxial diffraction.
The diffraction of beam or image is inevitable since its constituent plane wave components acquire different
phases during its propagation. The spatially dependent refractive index of the fields can be used to suppress 
or even reverse due to diffraction.
We use a suitable spatially dependent pump field to produce spatially dependent refractive index for the probe field.
For this purpose, we choose the transverse spatial profile of the pump beam as a Laguerre-Gaussian with charge $m$, denoted by LGP$_{m}$.
The profile of pump beam can be written as
\begin{align}
\label{profile_pump}
 &G(x,y,z)= G_{\rm 0} \frac{{w_{0}}}{{w_z}}\Bigg(\frac{r\sqrt{2}}{{w_z}}\Bigg)^m \exp\left[{\frac{ikr^2}{2R_z}-\frac{r^2 }{w_z^2}}\right]  \nonumber\\
          &\times \exp\left[ - i(m+1)\tan^{-1}\Bigg({\frac{z}{z_R}}\Bigg) + im\theta \right] \,,
\end{align}
where $G_0$ is an initial peak amplitude, $m$ is the azimuthal index. 
The beam width is defined as $w_{z}=w_{0}\sqrt{1 + \left({{z}/{{z_{R}}}}\right)^2}$, where 
$w_0$ is the beam waist at $z=0$, and $z_{R}={{\pi w_{0}^2 }}/{\lambda }$ is the Rayleigh length. 
The radial distance from the axis of the beam is given by $r=\sqrt {x^2 + y^2 }$.
Note that for the azimuthal index $m=0$, the Laguerre-Gaussian pump (LGP$_m)$ beam reduces to a Gaussian pump beam (GP$_0)$.
Figure {\ref{fig:Figure2}} shows the intensity distribution of the pump field against radial position $x$ at different lengths of the medium.
The LGP$_m$ beam exhibits a dark spot in the centre and a  bright profile in the annular region.
This makes intensity profile in contrast to GP$_0$ beam.
It is clearly shown in Fig.{\ref{fig:Figure2}}  that diffraction induced distortion of the pump beam profile is not severe even after 5 cm of propagation.
Therefore, phase modulation imposed on the probe beam due to the spatially varying pump beam
is effective throughout the length of the medium.
The probe beam possess a Gaussian profile 
\begin{align}
\label{profile_probe}
g(x,y)=&{g}_0 \:e^{-[\frac{ (x-a)^2+y^2}{w_p^{2}}]^f}\,,
\end{align}
at an entry face of the medium. 
The initial peak amplitude and the width of the probe field are denoted by $g_0$ and $w_p$ and 
$a$ is the initial location of the centre of the probe beam along the $x$ direction. 
We have chosen the initial intensity of the probe beam such that it gets absorbed inside 
the medium without pump and control fields. The integer values of $f$ decides the input profile of the 
probe beam - either a Gaussian $(f = 1)$ or a super-Gaussian $(f > 1)$.

\section{Results and Discussions}
\subsection{Spatial modulation of the probe field susceptibility}
\begin{figure}[t]
\subfigure[]
 {
 \centering
   \includegraphics[width=0.9\columnwidth,angle=0]{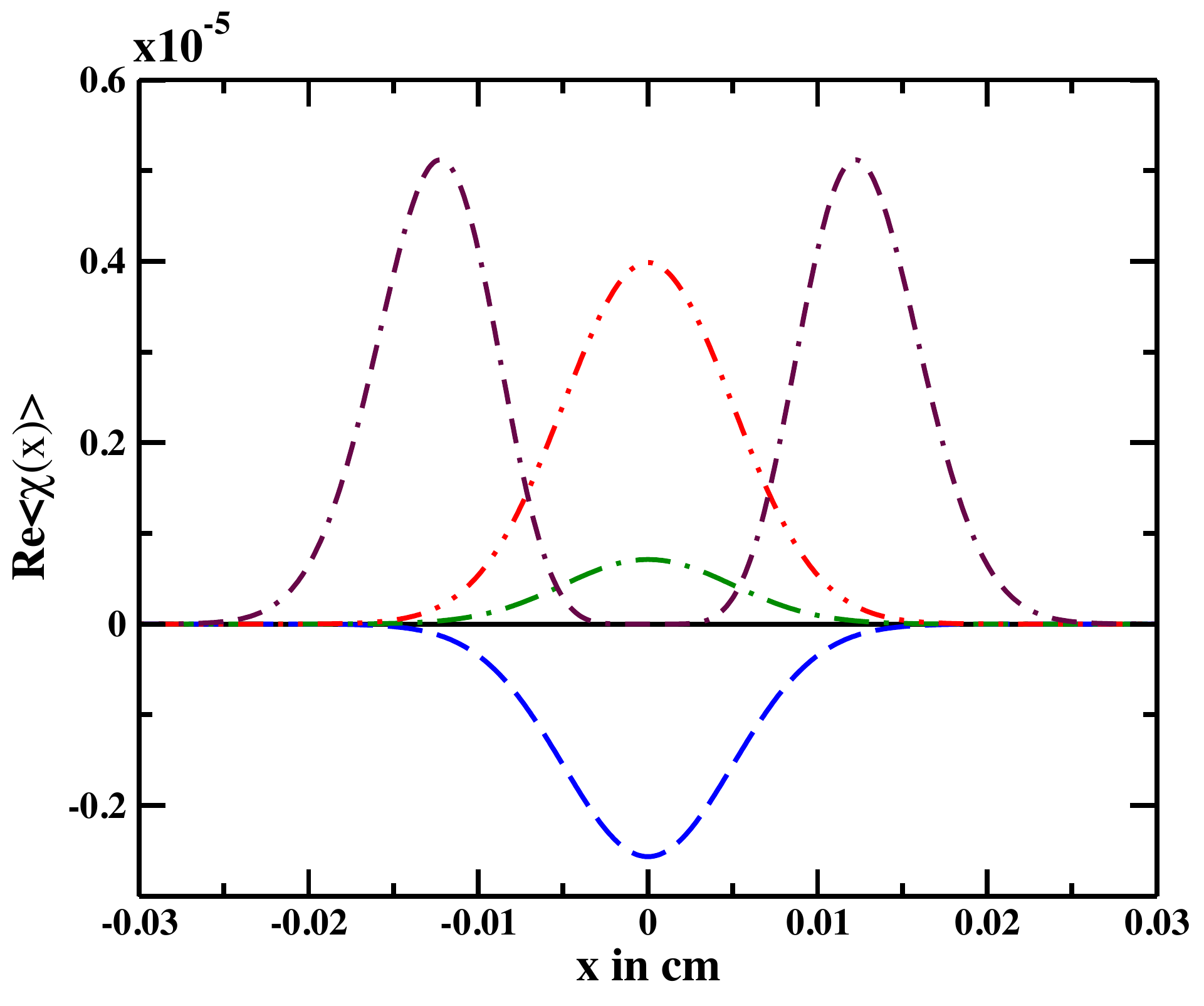}
   \label{fig:Figure3a}
 }
\subfigure[]
 {
 \centering
 \includegraphics[width=0.9\columnwidth,angle=0]{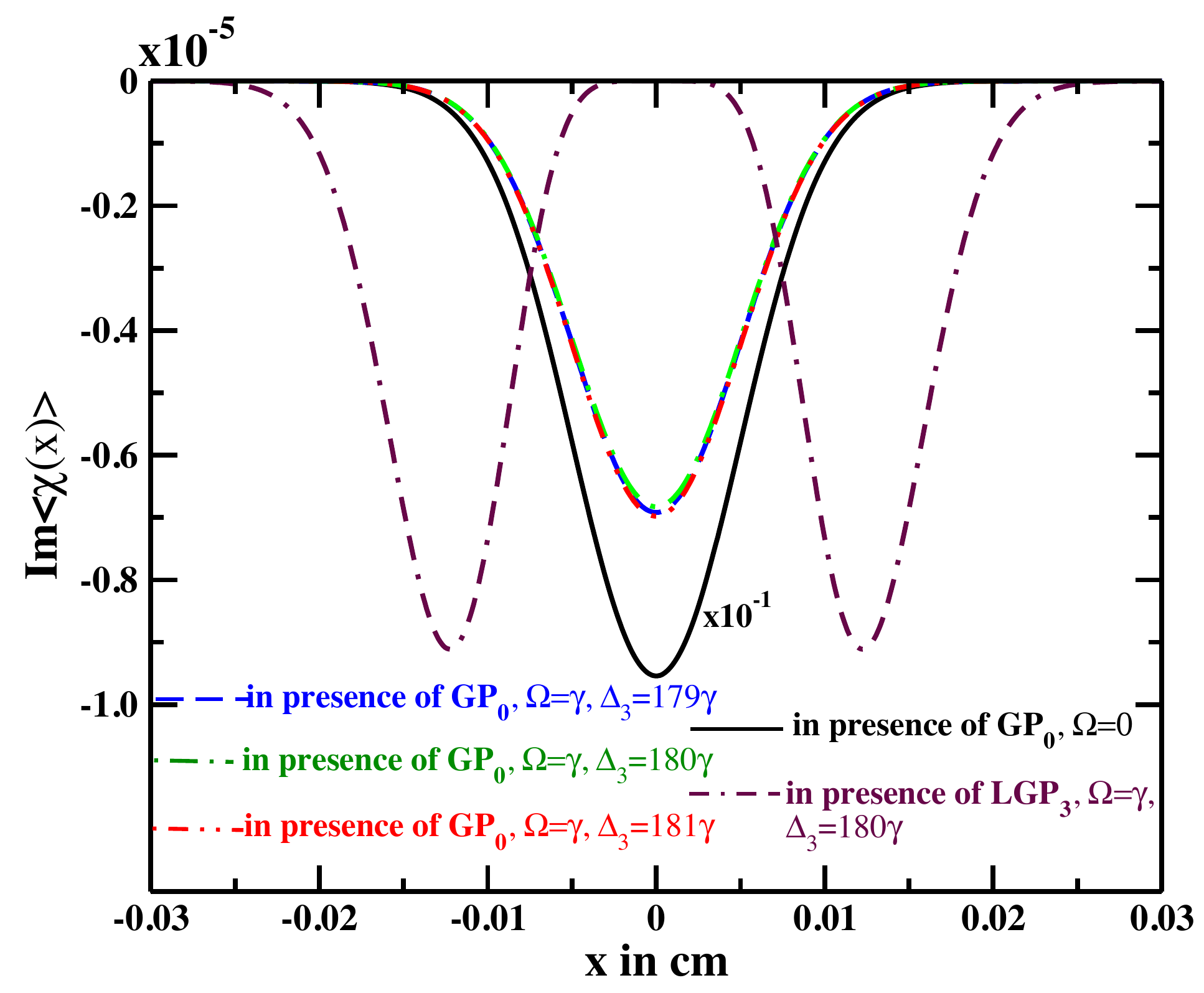}
   \label{fig:Figure3b}
 }
\caption{\label{fig:Figure3}(Color online)
\textcolor{black}{ Real and imaginary part of the averaged susceptibility is plotted against the transverse co-ordinate $x$ at $y=0$ plane.
The spatial probe gain profile (solid black line) is reduced by a factor of ten to visualise  it with  $\Omega(x,y)=\gamma$.
The common parameters are fixed as follows: single photon detuning of pump and probe fields $\DC=\DP=180{\gamma}$, Doppler width $\mathcal{D}=140\gamma$,  density $N=2.5\times10^{10}$ atoms/cm$^{3}$ and atomic coherence decay rate $\gamma_{_{c}}=0.01\gamma$. The other parameters are same as in Fig.\ref{fig:Figure2}
}}
\end{figure}

In order to elucidate the effect of position dependent characteristic of the pump field on the probe beam dynamics, we first numerically  explore the behaviour of velocity-averaged probe susceptibility under different detuning and intensity of the control field.
Fig.~\ref{fig:Figure3} shows the spatial variation of probe  dispersion and gain plotted against transverse axis $x$ at $y=0$ plane.
Here two different transverse profiles of pump beam namely Gaussian (GP$_{0})$ and Laugerre-Gaussian (LGP$_{3})$ have been used.
We begin with Gaussian pump beam and study the usefulness of uniform control field $\Omega$ on the spatially modulated probe susceptibility.
In absence of a control field, the position dependent refractive index of the probe is zero under two-photon Raman condition whereas the spatial gain profile of the probe field takes the shape of the pump beam profile.
The spatially dependent pump structure generates a probe gain profile which is  one of the key  components in realising the deflection of the probe beam if it is off-centered with respect to the pump beam.
In absence of the control field $\Omega=0$, the gain profile of the probe field is fifteen times larger than in case of control field $\Omega=\gamma$.  
This large gain can create modulation instability of system~\cite{agrawal1,agrawal2,Gisin}.
Therefore a controllable gain of the medium is required to avoid the modulation instability.
It is clear from Fig.~\ref{fig:Figure3} that the position dependent probe gain can be substantially suppressed by a uniform control field with  $\Omega(x,y)=\gamma$.
This restricted probe gain is accompanied by a Gaussian shaped spatial refractive index.
The gradient of the refractive index is dependent on the sign of the control field detuning.
At red control field detuning, the slope of the spatial refractive index attains maximum at the line centre and decreases gradually toward the wings.
Hence, a convex lens like refractive index  can be mimiced in the ARG medium for $\DM\ge\DP$.
On the contrary  the blueshifted the control field detuning  $\DM<\DP$ can  generate a concave refractive index profile onto the medium.
Therefore the  refractive index gradient  allows us to focus or defocus the probe beam towards the centre of the pump beam.
As a result  the probe field propagates through the gain window with narrowing or broadening, respectively.
Hence a control field can prepare a gain medium with suitable spatial refractive index for encompass the probe beam deflection to a great extent.

Next we consider higher order LGP$_3$ mode to investigate the spatial inhomogeneous character of $\langle\chi\rangle$ in the presence of uniform control beam.
The grey double dashed dotted line in Fig.~\ref{fig:Figure3a} and \ref{fig:Figure3b} shows the transverse variation of probe refractive index as well as gain, respectively.
The position dependent refractive index and gain both increase in the bright region whereas it decreases at the dark region of the doughnut shaped pump beam.
In other words, LGP$_3$ induces a diverging gradient index in the region $|r| \leq $ 0.005 cm whereas a converging gradient index exists in regions 0.005 cm$ \leq |r| \leq $ 0.02 cm of the medium.
Thus bright and dark regions of LGP$_3$ resembles a waveguide and anti waveguide structure inside the atomic medium.
Fig.\ref{fig:Figure3a} is also show that the waveguide and anti-waveguide features are accompanied with gain and absorption, respectively.
As a result, the probe beam is guided out from dark region and confined at bright region in the course of propagation inside the medium. 
Hence the shape of the pump beam profile can be efficiently transfered to the transmitted probe beam.
\begin{figure}[b!]
\subfigure[]
 {
 \centering
   \includegraphics[width=1.\columnwidth,angle=0]{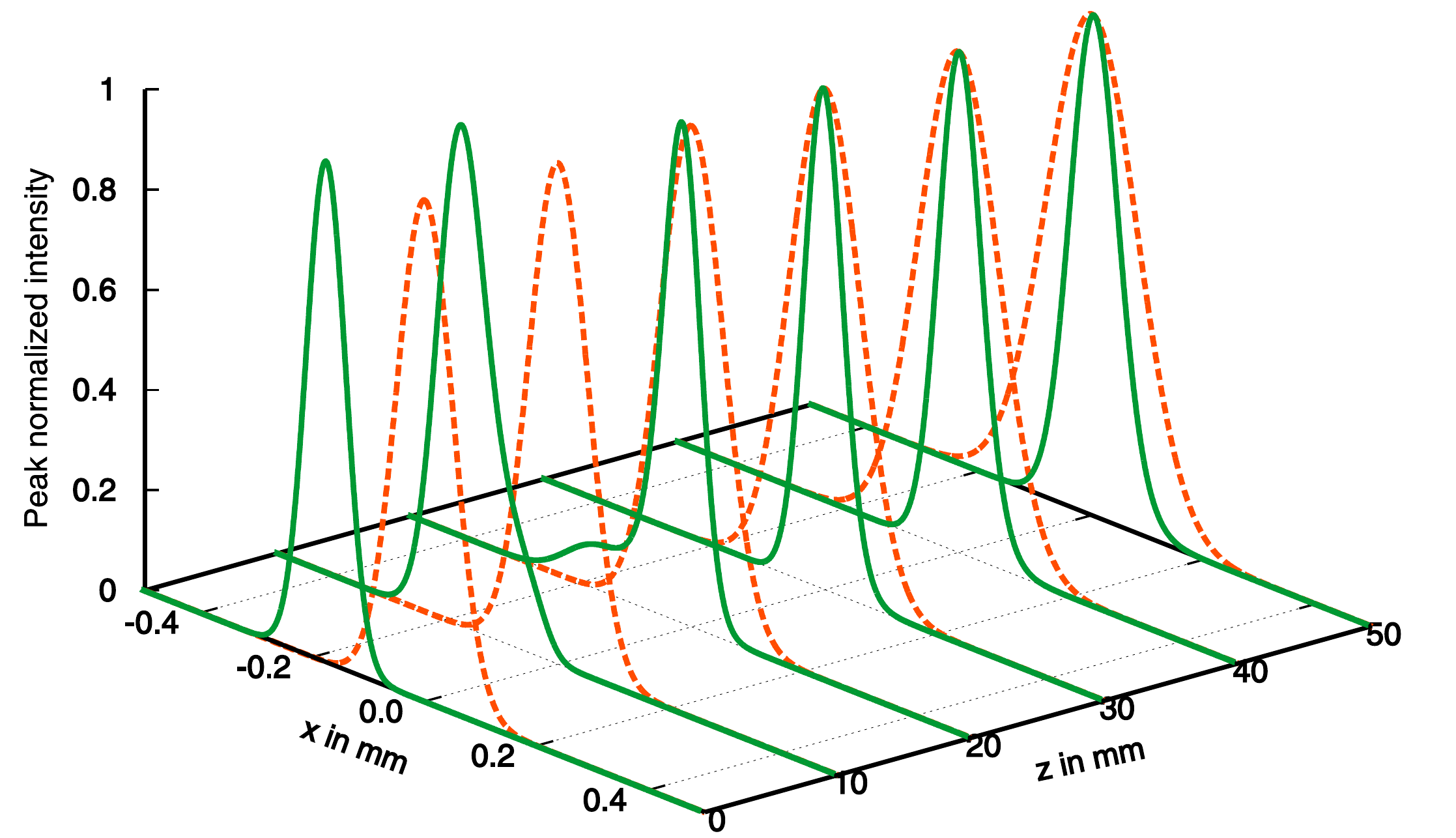}
   \label{fig:Figure4a}
 }
\subfigure[]
 {
 \centering
 \includegraphics[width=0.8\columnwidth,angle=0]{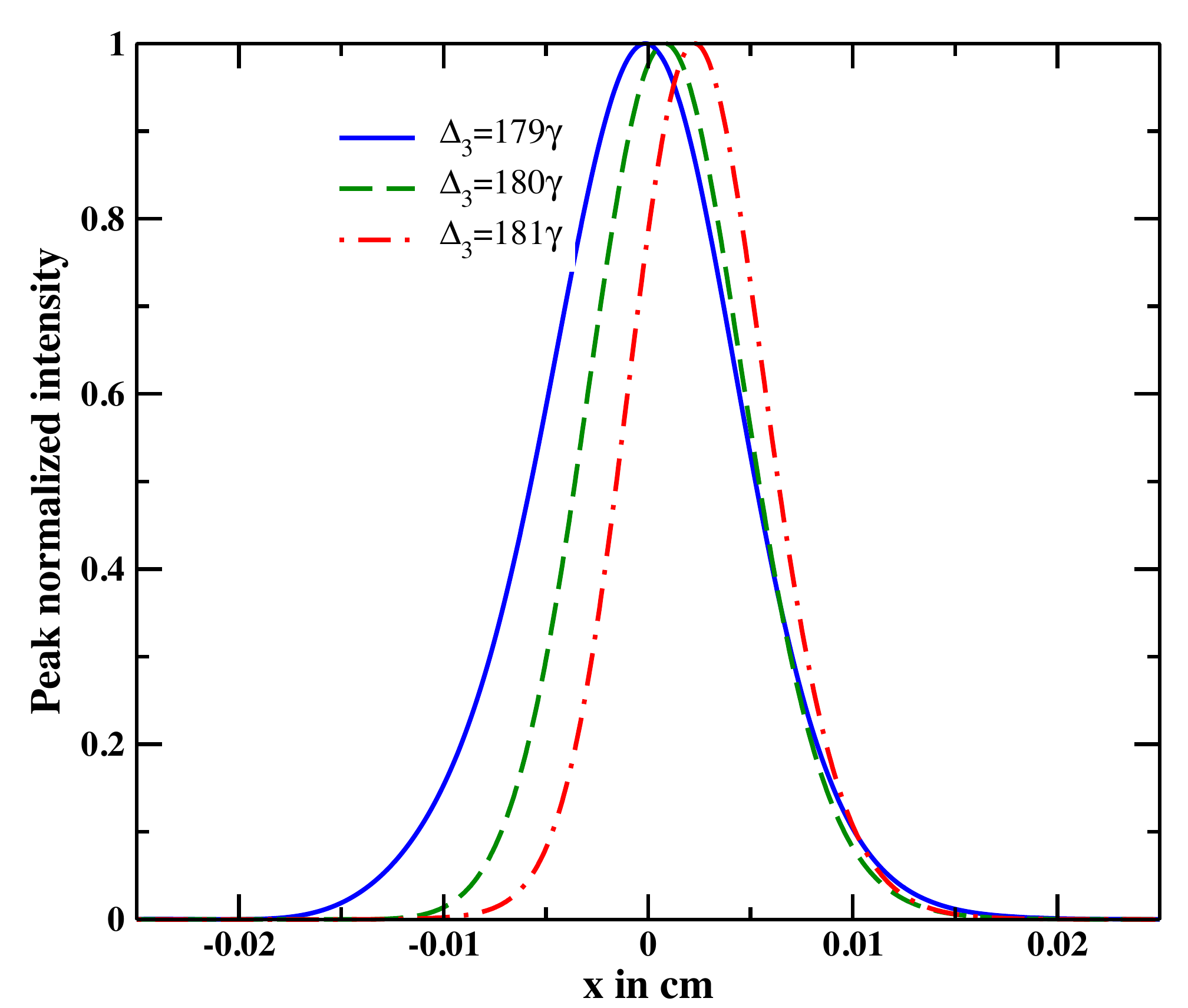}
   \label{fig:Figure4b}
 }
\caption{\label{fig:Figure4}(Color online)\textcolor{black}{ (a)
The transverse probe beam intensity is plotted at different propagation distances within the medium.
The initial amplitude, width and peak position of Gaussian probe beam are $g_{0}=10^{-3}\gamma$, ${w_p}=70~{\mu}$m and $a=1.7$~mm, respectively.
Single photon detuning of pump, probe and control fields are $\DC=\DP=\DM=180\gamma$.
(b) depict the transmitted probe beam width can be controllable by changing the detuning of control field at $z=4$ cm under two-photon Raman resonance condition $\DC=\DP=180\gamma$.
The other parameters are as in Fig.~\ref{fig:Figure3}.}}
\end{figure}
\subsection{ Numerical simulation of paraxial beams equations}
We have simulated numerically the propagation equations for pump ~(\ref{beam_eq_control}) and probe (\ref{beam_eq_probe}) beams by split step operator method~\cite{Shen} to demonstrate the spatial susceptibility  as well as  diffraction effects on the beam's propagation dynamics. 
\subsubsection{Optical beam steering} 
First, we study how the deflection of a probe beam can be controlled by a spatial dependence of the pump Rabi frequency. 
The shape and position of a probe beam is given by Eq~(\ref{profile_probe}) at the entry face of the medium. 
Fig.~\ref{fig:Figure4a} shows the spatial evolutions of the probe beam with $a=1.7$~mm and $w_p=70~\mu$m when the peak of pump beam is centered at the origin $(0,0)$ with $w_c=100\mu$m.
Initially the overlap area between the probe and pump beam is very negligible.
The overlap area is gradually increased due to the broadening of both the beams during propagation.
It is evident from this figure that after a propagation of one Rayleigh length, the probe beam progressively enters the pump region. 
The bright region of the pump beam tends to refract the probe beam into it  and subsequently enhances the probe beam amplitude.  
As a result, the probe beam is focused towards the high intensity region of the pump and remains confined there.   
Ii is noteworthy that the probe beam gains the initial shape of the pump beam and  retains this shape as it propagates along the $z$ axis. 
Similarly, if the peak position of the probe beam is  shifted along the positive $x$-direction then it can be dragged by the pump beam towards the pump line centre.

Fig.~\ref{fig:Figure4b} exhibits the effect of control field detuning onto the propagation dynamics of a probe pulse at $z=4$ cm.
It is seen that the deflected probe beam becomes narrower at redshifted detuning as compared to a blue-shifted detuning.
Therefore the sign of the detuning of the control field gives an additional flexibility to control the width of the deflected probe beam.
Thus the ARG medium not only acts as an effective beam deflector but also can act like a lens with a wide focal length tunability. 
\begin{figure}[t!]
\includegraphics[width=1.\columnwidth]{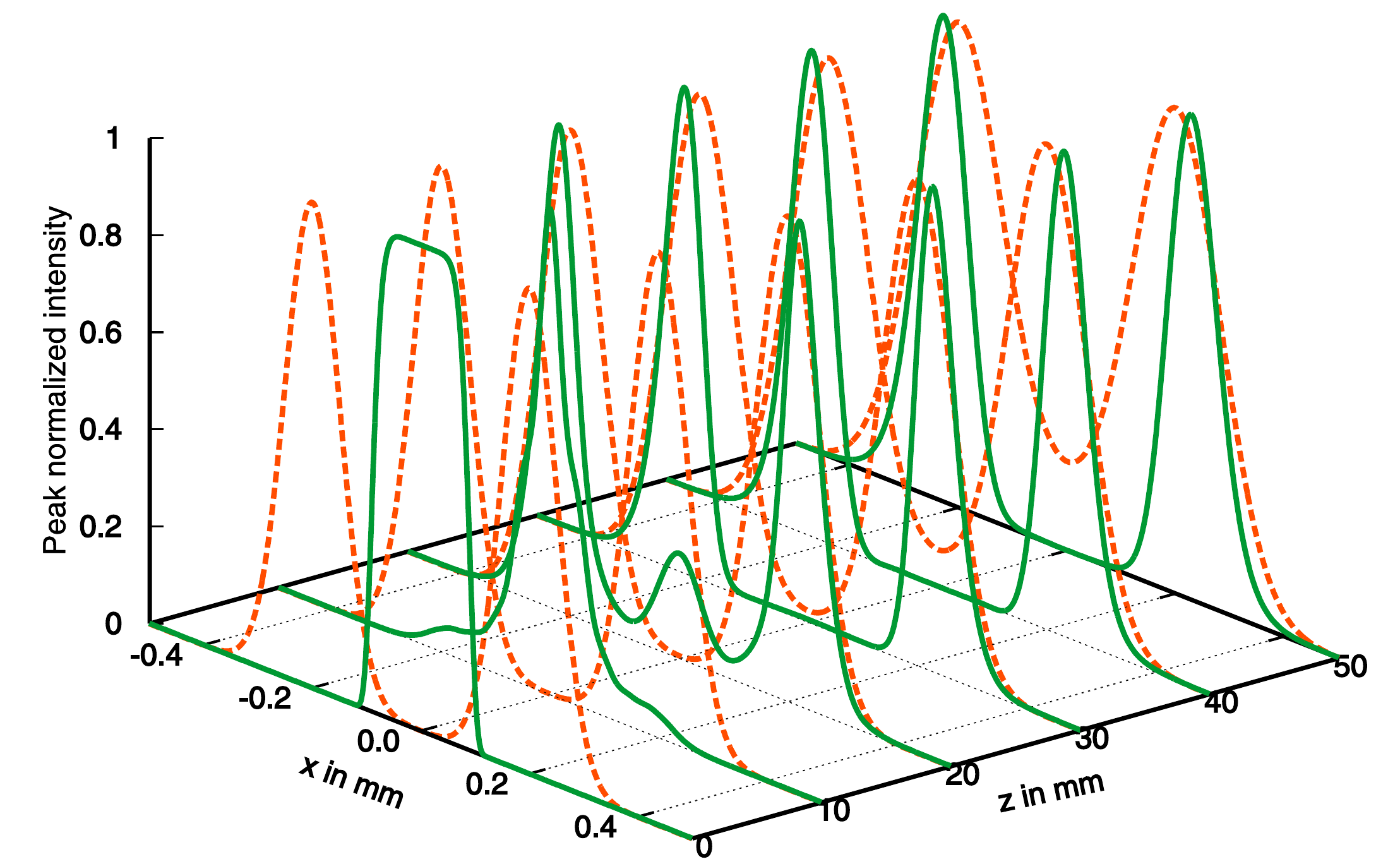}
\caption{\label{fig:Figure5} (Color online) 
\textcolor{black}{Propagation dynamics of single super-Gaussian probe beam in presence of double Gaussian pump beam.
The parameters are as in Fig.\ref{fig:Figure4a} except that the Gaussian probe beam is injected at centre $(0,0)$ with width $w_p=80\mu$m and the double Gaussian pump beam has width $100\mu$m.}}
\end{figure}
\subsubsection{Optical beam splitting}
Next, we demonstrate the spatial evolution of single super-Gaussian probe as well as double Gaussian pump beam with different propagation distance $z$.
At the entrance face of the medium, the probe beam is launched in the dark region of the double Gaussian pump beam as shown in Fig.~\ref{fig:Figure5}.
The position dependent  pump beam creates two gain peaks together with converging refractive index in the probe susceptibility which is similar to the grey double dashed dotted line in Fig.\ref{fig:Figure3}.
The gain and spatial inhomogeneity of the refractive index is accountable for this splitting of a single super-Gaussian probe beam into two Gaussian beams.
The converging lens effect in the intense regions of the pump leads to focusing of the cloned probe beam towards it.
The finesse of the transmitted probe beam can be defined as the ratio of the spacing between peaks to the width of peaks.
The transmitted probe beam width is reduced by a factor of 1.5  and the peaks separation are increased by 0.7 mm as compared to the initial shape of the pump beam. 
Hence the finesse of the cloned image has doubly enhanced as compared with initial pump image.
Noticeably from Fig.~\ref{fig:Figure5} the transmitted probe beam structure is preserved even though the pump beam suffers distortion due to diffraction.
\subsubsection{Optical beam cloning}

\begin{figure}[t]
\subfigure[]
 {
 \centering
   \includegraphics[width=0.9\columnwidth,angle=0]{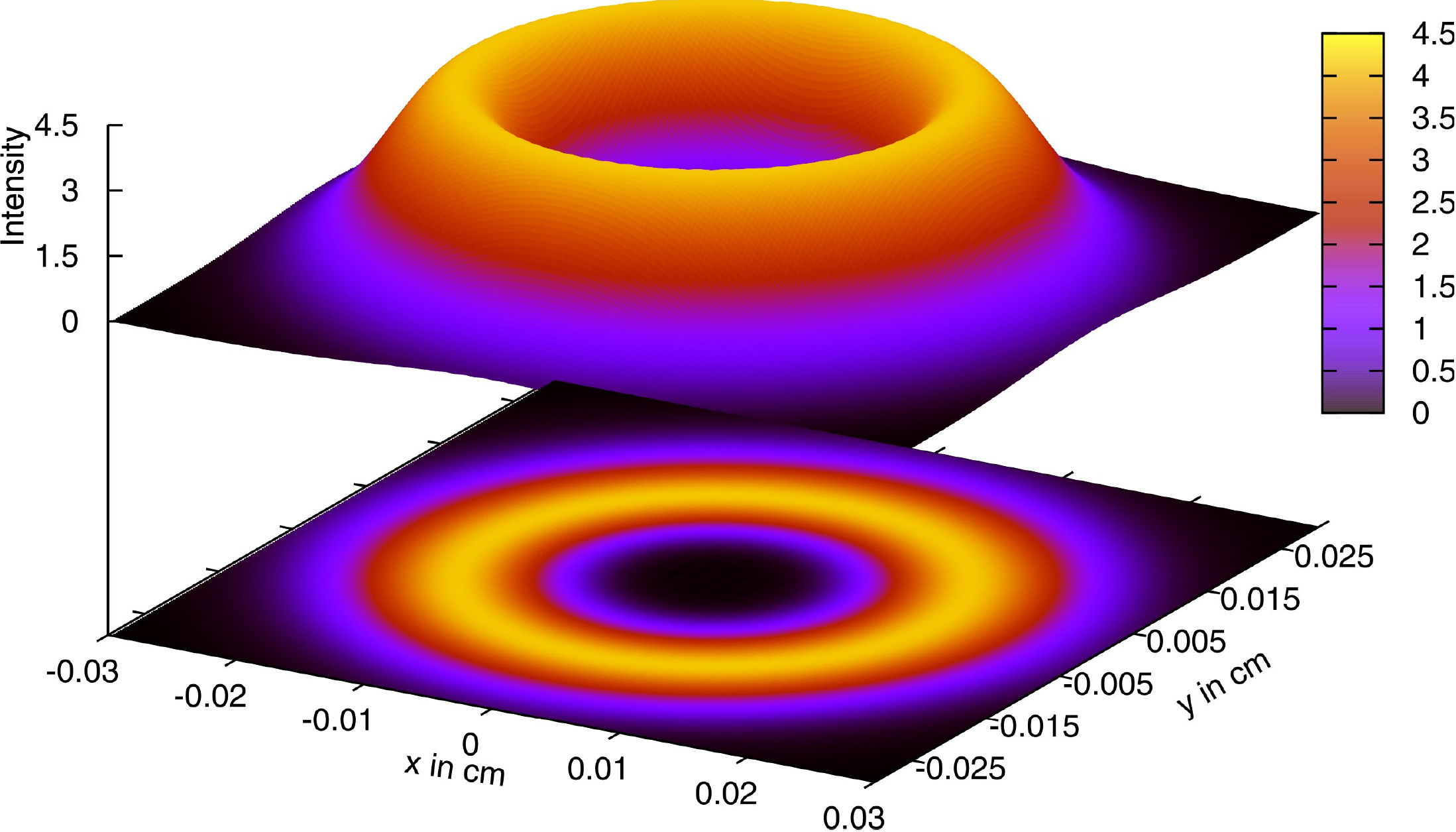}
   \label{fig:Figure6a}
 }
\subfigure[]
 {
\centering
 \includegraphics[width=0.9\columnwidth,angle=0]{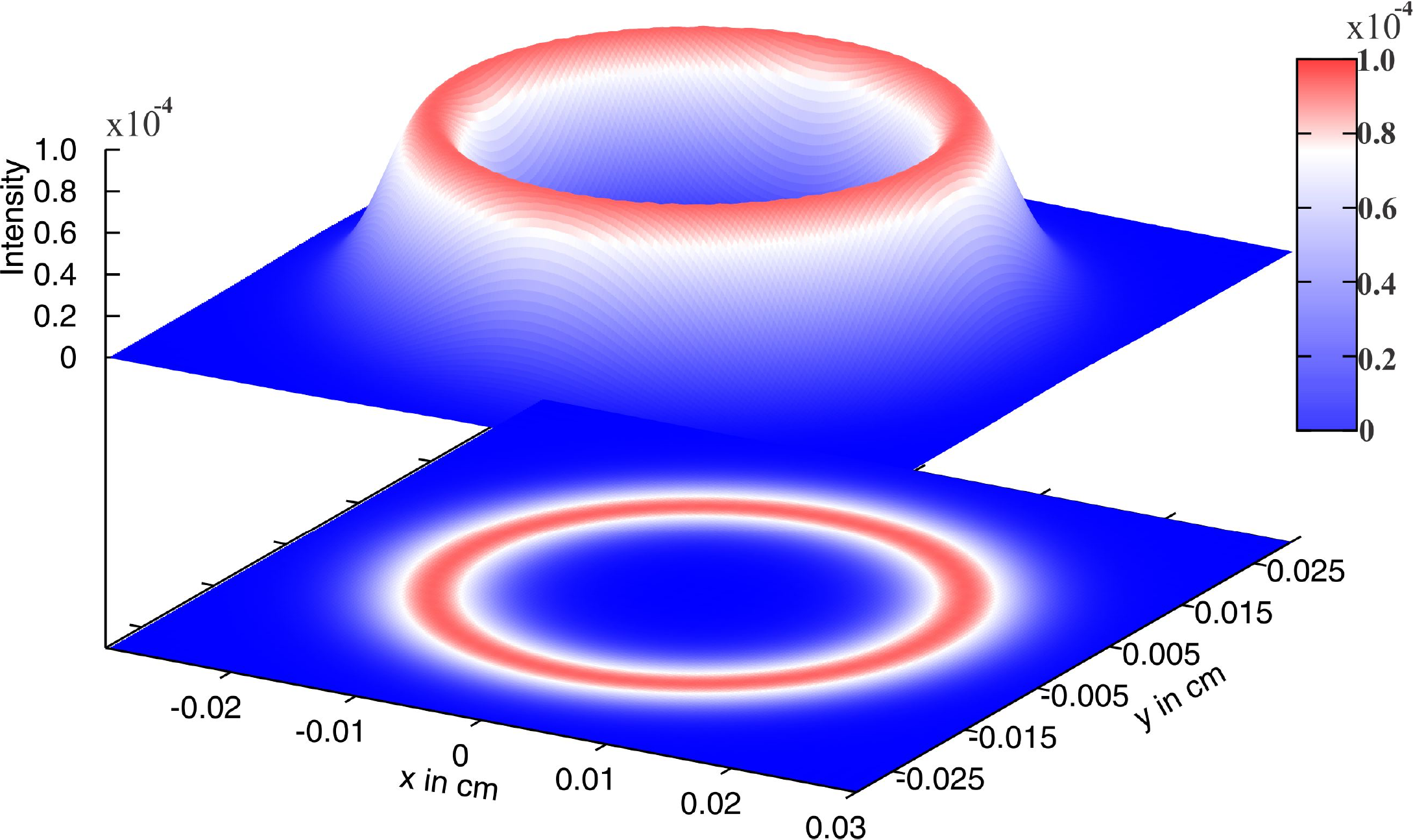}
   \label{fig:Figure6b}
 }
\caption{\label{fig:Figure6}(Color online)\textcolor{black}{Image transfer from doughnut-shaped pump structure  to the probe beam via anti-waveguiding mechanism.
In (a), the 3D intensity profile of pump beam at the output of 5-cm-long medium.
In (b), the cloned 3D probe intensity profile at the exit face of the rubidium vapour cell. 
The other parameters are same as in Fig.~\ref{fig:Figure4a} except atomic density $N=2.5\times10^{11}$ atoms/cm$^{3}$ and ground state atomic coherence decay rate $\gamma_{_{c}}=0.001\gamma$, $G_0=2.5\gamma$ and $\Omega=5\gamma$.}}
\end{figure}

In this section, we investigate the efficient transfer of images between two co-propagating orthogonal polarized optical beams.
We adopt all-optical anti-waveguide mechanism to clone the images from pump to probe beam.
An all-optical anti-waveguide structure can be formed inside the medium with use of LGP$_3$ beam which has zero intensity at the beam centre.
The dark and bright regions of  LGP$_3$ beam give rise to minimum and maximum refractive index  gradient on the probe susceptibility.
As a results a diverging and converging refractive index is formed in the core and cladding region of the anti-wave guide structure.
Thus an all-optical anti-waveguide for a probe beam is generated by the co-propagating doughnut-shaped strong pump beam.
In order to demonstrate the cloning mechanism in ARG medium, the centre of the dark region of the doughnut pump beam is the initial location of probe beam.
The  diverging refractive index gradient  and diffraction leads to the probe beam leaving the core region and slowly enter the high intensity regions of the pump beam.
Therefore each wing of the probe beam profile experiences gain and converging gradient of refractive index in the cladding region. 
Thus the probe energy is guided into the annular ring of the doughnut-shaped beam and leaving a zero intensity in the dark region.
Hence the transmitted probe beam profile acquires a doughnut-shaped profile as shown in Fig.~\ref{fig:Figure6}.
We have found that the transmitted structure of probe beam is twice as sharp compared to the LGP$_3$ beam structure.
The spatial evolution of probe beam at different propagation distances are similar to that in Fig.~\ref{fig:Figure5}. 
\begin{figure}[t!]
\includegraphics[width=1.\columnwidth]{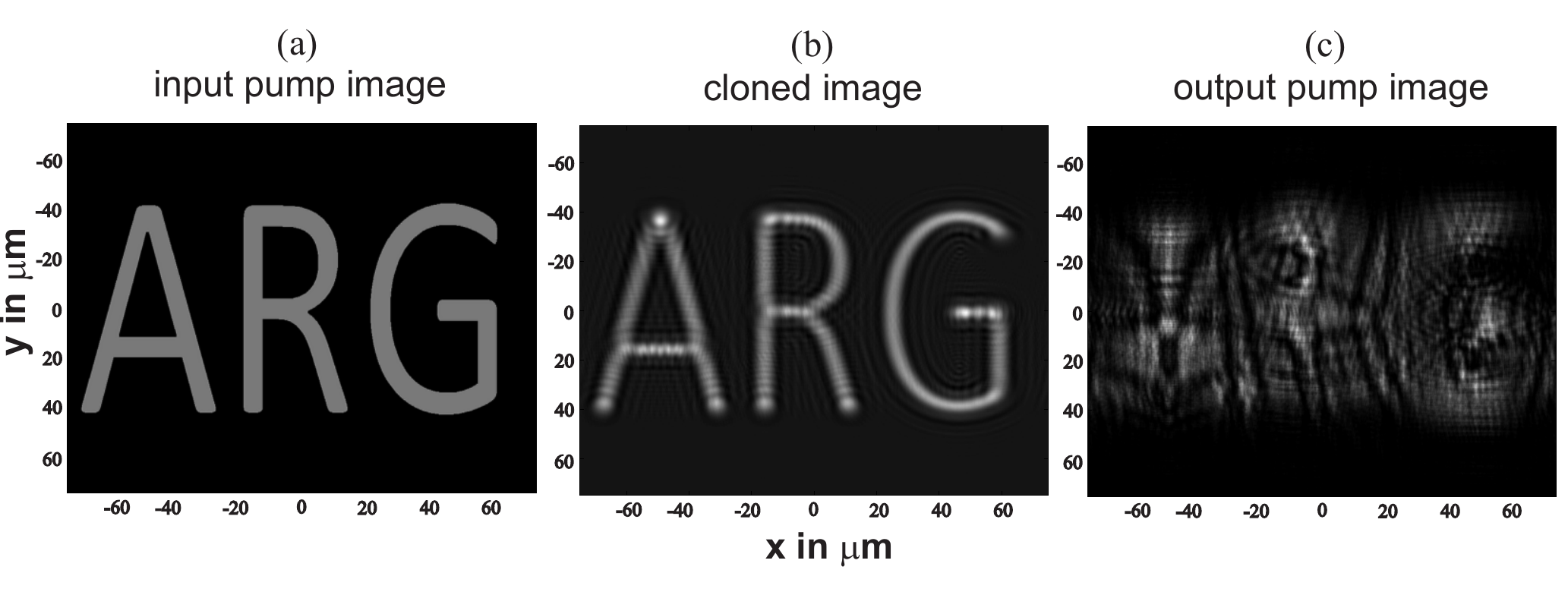}
\caption{\label{fig:Figure7}(Color online)\textcolor{black}{
(a) Three letters ``ARG" are imprinted on pump beam.
(b) The efficiently transferred image onto the probe beam after  2 cm length of propagation inside the atomic medium. 
(c) The transmitted pump beam image which is completely blurred at the exit face of medium.
The parameters are same as in Fig.\ref{fig:Figure6}.}}
\end{figure}
\subsubsection{Arbitrary image cloning}
Fig.~\ref{fig:Figure7} shows the  cloning of an arbitrary images and its diffraction effects through an ARG medium. 
In order to elucidate the arbitrary image cloning, we consider the probe beam as a plane wave whereas the pump beam carries complex image such as three letters ``ARG" structured at the entrance face of the medium. 
The two-dimensional transverse profile of the pump beam creates gain for the probe beam wherever two-photon Raman condition is satisfied.
Hence the  transverse pattern of the pump beam can be efficiently transferred  to the probe beam.
The cloned probe beam also experiences focusing effects at the high intensity regions of the pump beam.
Thus the transmitted probe beam has better resolution than the original pump beam images as can be seen in Fig.~\ref{fig:Figure7}b. 
Fig.~\ref{fig:Figure7}c illustrates that the diffraction induced distortion severely affects the pump beam images and is completely distorted after a propagation of 2 cm. 
\section{Conclusion}
In conclusion, we have studied diffractionless steering, splitting and cloning of an optical beam in a Doppler broadened 
four level $\mathcal{N}$-type Raman gain medium using a spatially inhomogeneous pump beam.
The spatial pump beam profile gives rise to transverse modulation in the refractive index and gain for the probe beam.
The modulated refractive index along with gain can optically form anti-waveguide structure inside the medium.
The properties of anti-waveguide structure such as refractive index and gain can be controlled by the application of 
control field which lead to steering of the probe beam very efficiently.
We further demonstrated that a single probe beam can be split into two Gaussian modes when it is injected at the centre between two Gaussian modes of pump beam. 
We found that the probe beam profile has acquired the shape of the pump beam and propagates without usual diffraction.
We next show that the transfer of doughnut-shaped pump image onto a low power Gaussian-shaped probe beam can be possible with high finesse.  
Finally, by numerical simulations we have established that an arbitrary image with three letters ``ARG" imprinted on pump beam can be cloned on to the transmission profile of the probe. 
The finesse of cloned image has increased twice as compared to the initial resolution of pump images. 
Thus this scheme might be useful in optical switching, optical lithography, and optical imaging processing.
\begin{acknowledgements}
{TND acknowledges Science and Engineering Board of India for financial support (SR/S2/LOP-0033/2010).}
\end{acknowledgements}
\appendix
\section{\label{app-A}Expressions of probe susceptibility}
The related $(12)$th-order contributions for $\rho_{_{32}}$are obtained as
\begin{align}
{\rho}_{_{32}}^{^{(12)}} &= \frac{ \ii ({\rho}_{_{22}}^{^{(6)}} - {\rho}_{_{33}}^{^{(6)}}) + \ii{\rho}_{_{12}}^{^{(9)}} - \ii{\Omega}{\rho}_{_{34}}^{^{(12)}} } {\left[\gamma_{_{32}} - \ii \DP\right]}\\
{\rho}_{_{34}}^{^{(12)}} &= \frac{ \ii {\rho}_{_{14}}^{^{(9)}} + \ii{\rho}_{_{24}}^{^{(6)}} - \ii{\Omega^*}{\rho}_{_{32}}^{^{(12)}}} {\left[ \Gamma_{_{34}} - \ii (\DP-\DM)\right]} \\
{\rho}_{_{12}}^{^{(9)}} &=  \frac{\ii {\rho}_{_{32}}^{^{(3)}}  -\ii{\rho}_{_{13}}^{^{(2)}} - \ii{\Omega}{\rho}_{_{14}}^{^{(9)}}} {\left[ \gamma_{_{c}} + \ii \DR\right]} \\
{\rho}_{_{14}}^{^{(9)}} &=  \frac{ i{\rho}_{_{34}}^{^{(1)}} -\ii {\Omega^{*}}{\rho}_{_{12}}^{^{(9)}} }{\left[ \Gamma_{41} + \ii (\DR+\DM)\right]}\\
{\rho}_{_{33}}^{^{(6)}} &=  \frac{ i ({\rho}_{_{13}}^{^{(2)}} - {\rho}_{_{31}}^{^{(1)}}) } { \left[ \gamma_{_{13}} + \gamma_{_{23}} \right]}\\
{\rho}_{_{24}}^{^{(6)}}&= {\rho}_{_{44}}^{^{(6)}}={\rho}_{_{22}}^{^{(6)}}=0\\
{\rho}_{_{13}}^{^{(2)}} &= {\rho}_{_{31}}^{^{(1)}}= \frac{ -\ii }{\left[ \Gamma_{_{13}} + \ii \DC\right]}  \\
{\rho}_{_{32}}^{^{(3)}} &= {\rho}_{_{34}}^{^{(1)}}=0  
\end{align}
\begin{widetext}
\begin{align}
\label{raman_chi}
{\rho}_{_{32}}&=  {A}
\left[\frac{{2\Gamma_{_{31}}[\Gamma_{_{34}} - \ii (\DP-\DM)]}}{(\gamma_{_{13}}+\gamma_{_{23}})(\Gamma_{_{31}}^2+\DC^2)}
+\frac{{[\Gamma_{_{34} }- \ii (\DP-\DM)][\Gamma_{_{41}} + \ii (\DR+\DM)]}
-|\Omega|^2}{(\Gamma_{_{31}} + \ii \DC)\left[(\gamma_{_{c}} + \ii \DR)(\Gamma_{_{41}} + \ii (\DR+\DM))+|\Omega|^2\right]}\right]
\end{align}
\end{widetext} 
with
\begin{align}
{A} &=  \frac{-\ii g |G|^2} {({\gamma_{32} - \ii \Delta_2})\{\gamma_{34} - \ii (\Delta_2-\Delta_3)\}+{|\Omega|^2}}
\end{align}

\end{document}